\documentclass[11pt]{article}
\pdfoutput=1
\usepackage{jcappub}
\usepackage{bm}
\usepackage{amsmath}
\usepackage{verbatim}

\def\ei{\end{itemize}}
\def\be{\begin{equation}}
\def\ee{\end{equation}}
\newcommand{\bea}{\begin{eqnarray}}
\newcommand{\eea}{\end{eqnarray}}

\title{
Inflationary magneto-(non)genesis, increasing kinetic couplings, and the strong coupling problem
}
\author[a,b]{Hossein Bazrafshan Moghaddam,}
\author[a,c]{Evan McDonough,}
\author[a]{Ryo Namba,}
\author[a]{and Robert H.~Brandenberger~}
\affiliation[a]{Department of Physics, McGill University, Montr\'{e}al, QC, H3A 2T8, Canada}
\affiliation[b]{Department of Physics, Ferdowsi University of Mashhad, P.O.Box 1436, Mashhad, Iran}
\affiliation[c]{Department of Physics, Brown University, Providence, RI, USA. 02903}
\emailAdd{bazrafshan@physics.mcgill.ca}
\emailAdd{evanmc@physics.mcgill.ca}
\emailAdd{namba@physics.mcgill.ca}
\emailAdd{rhb@hep.physics.mcgill.ca}

\abstract{
We study the generation of magnetic fields during inflation making use of a coupling of
the inflaton and moduli fields to electromagnetism via the photon kinetic term, and assuming that
the coupling is an {\it increasing} function of time. We
demonstrate that the strong coupling problem of inflationary magnetogenesis can be avoided 
by incorporating the destabilization of moduli fields after inflation.
The magnetic field always dominates over the electric one, and thus the severe constraints 
on the latter from backreaction, which are the demanding obstacles in the case of a decreasing 
coupling function, do not apply to the current scenario.
However, we show that this loophole to the strong coupling problem comes at a price: 
the normalization of the amplitude of magnetic fields is determined by this coupling term and is therefore suppressed by a large factor after the moduli destabilization completes.
From this we conclude that there is no self-consistent and generic realization of primordial 
magnetogenesis producing scale-invariant fields in the case of an increasing kinetic coupling.}

\keywords{inflation, reheating, primordial magnetogenesis}
\arxivnumber{1707.05820}

\begin{document}

\maketitle

%
%
%
\section{Introduction}

Observations of blazars \cite{Neronov:1900zz, Tavecchio:2010mk, Dolag:2010ni, Essey:2010nd, Taylor:2011bn, Takahashi:2013lba, Finke:2013tyq, Chen:2014rsa} hint at the existence of 
magnetic fields in extragalactic regions, where astrophysical processes are unlikely to be responsible for their generation. They set the lower bound on the effective strength of the fieds as \cite{Taylor:2011bn, Neronov:2009gh},%
\footnote{Depending on the specific assumptions in the analysis, such as the intrinsic energy spectrum and time variation of the source, the lower bound varies between $\sim 10^{-17} \, {\rm G}$ and $\sim 10^{-15} \, {\rm G}$.}
\begin{equation}
B_{\rm eff} \gtrsim 10^{-15} \, {\rm G} \; ,
\label{Beff}
\end{equation}
where
\begin{equation}
B_{\rm eff} \equiv B \times 
\begin{cases}
\displaystyle
\sqrt{\frac{\lambda}{1 \, {\rm Mpc}}} & , \quad (\lambda \lesssim 1 \, {\rm Mpc}) \\
1  & , \quad (\lambda \gtrsim 1 \, {\rm Mpc}) 
\end{cases}
\end{equation}
and $B$ is the current amplitude of the magnetic fields with coherence length $\lambda$. The dividing line at $\lambda \sim 1 \, {\rm Mpc}$ is fixed by the mean free path of the electrons in the intergalactic media, those pair-produced by the scatterings of blazar TeV $\gamma$ rays with extragalactic background light \cite{pairproduction,Durrer:2013pga}.

A compelling possibility to generate such extragalactic magnetic fields is to attribute their origin to cosmology (for recent reviews, see \cite{Kandus:2010nw, Durrer:2013pga, Subramanian:2015lua}).
For example, collision of bubbles created at phase transitions in the early universe \cite{Vachaspati:1991nm, Sigl:1996dm} or anomalous global strings from QCD coupled to electromagnetism \cite{Brandenberger:1998ew} can produce magnetic fields, while a concrete quantitative model aiming for blazar observations is yet to be established. Also, the motion of charged particles in the plasma can induce magnetic fields at some cosmological scales in second-order perturbation theory \cite{Matarrese:2004kq, Saga:2015bna, Fidler:2015kkt}, though this mechanism has been shown to be unable to achieve the production at the observed level.
Generation of magnetic fields in non-singular cosmologies has also been considered, such as in the pre-big bang scenario \cite{Gasperini:1995dh} and in a bouncing universe \cite{Battefeld:2004cd,Salim:2006nw, Membiela:2013cea, Sriramkumar:2015yza, Qian:2016lbf, Chowdhury:2016aet}.

A large body of work has explored the possibility that magnetogenesis can be incorporated into the context of inflationary cosmology \cite{Turner:1987bw, Ratra:1991bn, Garretson:1992vt, Finelli:2000sh, Davis:2000zp, Bamba:2003av, Anber:2006xt, Martin:2007ue, Demozzi:2009fu, Kanno:2009ei, Durrer:2010mq, Barnaby:2012tk, Fujita:2012rb, Ferreira:2013sqa, Fujita:2013pgp, Fujita:2014sna, Kobayashi:2014sga, Ferreira:2014hma, Kobayashi:2014zza, Caprini:2014mja, Tasinato:2014fia, Fujita:2015iga, Campanelli:2015jfa, Green:2015fss, Domenech:2015zzi, Campanelli:2015lqz, Fujita:2016qab, Adshead:2016iae, Mukohyama:2016npi, Vilchinskii:2017qul, Markkanen:2017kmy}. The most studied candidate for inflationary magenetogenesis is the kinetic coupling model originally proposed by Ratra \cite{Ratra:1991bn}, which utilizes a coupling through the electromagnetic kinetic term of the form
\be
{\cal L}_{\rm Ratra \; model} = -  \frac{I^2(\varphi)}{4} \, F_{\mu \nu} F^{\mu \nu} \; ,
\label{Ratra}
\ee
with $I^2(\varphi) = {\rm e}^{\alpha \varphi}$ in the original work, $\alpha$ being a constant, where $\varphi$ is the inflaton and $F_{\mu \nu}$ is the photon field-strength tensor. During inflation, the dynamics of $I(\varphi)$ leads to the production of gauge field quanta, while today, the value of $I(\varphi)$ is unity, in order to match with the standard electromagnetism.%
\footnote{This constant normalization is degenerate with that of the photon field $A_\mu$, and thus it is matter of our choice. We take the normalization $I = 1$ at present throughout this paper.}
Another class of well-studied models employs a pseudo-scalar field that has coupling to electromagnetism, uniquely allowed by symmetries \cite{Garretson:1992vt, Finelli:2000sh, Anber:2006xt, Durrer:2010mq, Caprini:2014mja, Fujita:2015iga, Adshead:2016iae}, where produced magnetic fields may receive substantial support due to the inverse cascade process of the turbulent plasma in the early unviverse (e.g.~\cite{Brandenburg:1996fc, Banerjee:2004df, Campanelli:2007tc, Boyarsky:2011uy, Tashiro:2012mf, Saveliev:2013uva, Boyarsky:2015faa, Pavlovic:2016gac}).

In models of the type \eqref{Ratra}, the time evolution of $I$ is not fixed {\it a priori} and is controlled by its functional form and by the motion of $\varphi$.
A monotonic time dependence during inflation, starting from an initially large value and ending at unity, has often been considered as simple cases, and in particular,
a scale-invariant spectrum of magnetic fields is realized via a time evolution during inflation given by \cite{Martin:2007ue}
\be
I(\varphi) \propto a^{-3}(t) .
\ee
However, as we review in Sec.~\ref{sec:constraints}, models with decreasing $I(\varphi)$ store the majority of their electromagnetic energy in {\it electric} fields, and it is not possible to realize sufficient production of {\it magnetic} fields without significant back-reaction of the electric fields on the inflationary dynamics \cite{Demozzi:2009fu, Kanno:2009ei, Fujita:2012rb} and/or on the cosmic microwave background (CMB) observables \cite{Barnaby:2012tk, Fujita:2013pgp, Fujita:2014sna, Ferreira:2014hma}. Thus these models have a difficulty sometimes called the ``back-reaction problem.''

An alternative to the above is to consider choices of $I(\varphi)$ that are \emph{increasing} during inflation. A scale-invariant spectrum of magnetic fields is realized in this context by \cite{Martin:2007ue}
\be
I(\varphi) \propto a^{2}(t) 
\ee 
during inflation. 
In this case the electromagnetic energy density is dominated by its {\it magnetic} component, and therefore the stringent ``back-reaction problem'' in the case of decreasing $I$, mentioned above, appears to be irrelevant.
However, these models are plagued by their own difficulty: The effective electric charge, given by
\be
e_{\rm eff} = \frac{e}{I} ,
\ee
becomes large when $I$ is small. Since $I=1$ today and is increasing during inflation, the value of $I(\varphi)$ during inflation (at the time when primordial magnetic fields are produced) is necessarily much smaller than one, if one assumes a simple monotonic function in time. The electromagnetic coupling is thus large, and the system is strongly coupled. In this case, correlators of electromagnetic fields cannot be reliably computed in perturbation theory, and the model is said to suffer a ``strong coupling problem'' \cite{Demozzi:2009fu}.%
\footnote{See also \cite{Kobayashi:2014zza, Hayashinaka:2016qqn, Hayashinaka:2016dnt, Geng:2017zad} for a variant of the strong coupling problem arising from the Schwinger effect.}

Some authors have advocated the consideration of the above strong coupling regime. These models easily avoid the backreaction and CMB constraints, but have been considered to require significant modification of the standard model sector \cite{Tasinato:2014fia} or violation of gauge invariance \cite{Domenech:2015zzi}. Other authors have taken a phenomenological approach, without a model implementation in the language of field theory, and considered a changing $I$, which may now be a function of non-inflaton fields, even after inflation \cite{Kobayashi:2014sga,Tsagas:2014wba, Tsagas:2015mza,Fujita:2016qab}. These models may be able to produce a sufficient amount of magnetic fields but require rather dedicated, and possibly fine-tuned, model building. It has also been proposed that the above problems are resolved by tuning the energy scale of inflation to be near the BBN scale \cite{Fujita:2014sna, Ferreira:2014hma}. Beyond these attempts, there is no known solution to the above problems.

In this work we point out a loop-hole in the above arguments allowing us to evade the strong-coupling problem, starting from the realization that the overall normalization of $I$ may depend on fields which are non-dynamical during inflation (e.g. stabilized string theory moduli), whose vacuum state during inflation may differ from that after inflation. 
We show that the strong coupling problem can be avoided in the ``increasing-$I$'' models by utilizing non-trivial dynamics of moduli fields near the end of inflation, or during (p)reheating, which leads to a jump in the value of $I$ from a large value at the end of inflation to $1$. 
To this end, we promote the function $I(\varphi)$ in \eqref{Ratra} to one not only of the inflaton $\varphi$ but also of a mudulus  field $\chi$. Specifically, we consider a model with the electromagnetic kinetic term
\begin{equation}
{\cal L}_{\rm our \; model} = - \frac{I^2(\varphi , \chi)}{4} \, F_{\mu\nu} F^{\mu\nu} \; ,
\end{equation}
where the evolution of $I$ is controlled by $\varphi$ until the end of inflation and is subsequently taken over by $\chi$.
We demonstrate that this can occur given only minimal assumptions on the theory and thus is a fairly general phenomenon. 

However, we show in this paper that this approach ultimately
can only produce a very suppressed amplitude of scale-invariant magnetic fields, not even close to the value \eqref{Beff}. This is precisely because of the normalization controlled by $\chi$ in our proposed mechanism. Since $I$ increases during inflation, magnetic fields are produced substantially more than the electric counterpart. In this case, the scale-invariant part of their amplitude evolves simply as $B \propto I/a^2$, and due to the drop of the value of $I$ from a huge value at the end of inflation to unity at a later time, it is subject to this large suppression, additional to the usual dilution factor $1/a^2$. This result is inevitable for the generation of large-scale magnetic fields, whenever electric fields are subdominant $\vert \vec{E} \vert \ll \vert \vec{B}\vert$ on cosmological scales, (otherwise the ``back-reaction problem'' mentioned above would again become prominent).
We are thus forced to conclude that the price of evading the strong-coupling problem with an increasing $I$ is to erase the scale-invariant magnetic fields and therefore this does not provide a viable mechanism for magnetogenesis at large scales.%
\footnote{See also \cite{Sharma:2017eps} (released shortly following this paper), where a phenomenological form of $I(a)$ is chosen (in a similar spirit to \cite{Kobayashi:2014sga,Fujita:2016qab}) that increases during inflation and decreases after, with the large-scale magnetic fields originating in the decreasing phase. The magnetic field spectrum in this case is blue-tilted, and similar to \cite{Fujita:2014sna,Ferreira:2014hma}, requires low-scale inflation to have observable magnetic fields on Mpc scales.}.

Our paper is organized as follows. In Sec.~\ref{sec:constraints}, we review the basic mechanism of electromagnetic production in the $I^2 F^2$ models together with the strong coupling issue, and summarize the known constraints from backreaction and CMB observations. Sec.~\ref{sec:increasing} introduces our mechanism to avoid the strong-coupling problem while evading these constraints. In Sec.~\ref{subsec:problem} we show that only a negligible magnitude of scale-invariant magnetic fields remains after $I\to 1$ is achieved. Then in Sec.~\ref{subsec:jumpmodel} we introduce
a concrete model to realize the mechanism of Sec.~\ref{sec:increasing}.
Sec.~\ref{sec:conclustion} is devoted for our conclusion. In Appendix \ref{app:preheating}, we discuss the parametric resonance of perturbations during preheating and demonstrate that this does not circumvent the suppression coming from $I$.

\section{Constraints on inflationary magnetogenesis}
\label{sec:constraints}

The most studied model of magnetogenesis in the primordial universe takes the electromagnetic (EM) Lagrangian of the form
\begin{equation}
{\cal L}_{\rm EM} = - \frac{I^2}{4} \, F_{\mu\nu} F^{\mu\nu} \; ,
\label{Lag-EM}
\end{equation}
where $F_{\mu\nu}$ is the field-strength tensor of the EM field $A_\mu$, and $I$ is a function of other fields that controls the time dependence of the kinetic coefficient. The variation of \eqref{Lag-EM} with respect to $A_\mu$ gives the equation of motion
\begin{equation}
\nabla_\nu \left( I^2 F^{\mu\nu} \right) = 0 \; ,
\label{EQ_A}
\end{equation}
where $\nabla_\mu$ is the covariant derivative associated with the metric $g_{\mu\nu}$. Assuming a vanishing vacuum expectation value (vev) of $A_\mu$,
and taking the Coulomb gauge $A_0 = \partial_i A_i = 0$, we decompose $A_i$ as
\begin{equation}
{\bar I} A_i(\tau, \vec x) = \int \frac{d^3k}{(2\pi)^{3/2}} \sum_{P} \left[ {\rm e}^{i \vec x \cdot \vec k} \, e_i^{P} \big( \hat k \big) \, V_P (\tau , k) \, a_{P, \vec k} + {\rm h.c.} \right] \; ,
\label{A-FT}
\end{equation}
where $\tau$ is the conformal time, $e_i^P$ are the circular polarization vectors, and $\bar I \equiv  \sqrt{\langle I^2 \rangle}$ is understood as the background, time-dependent quantity.
Then the equation of motion for the mode function $V_P$ reads
\begin{equation}
\partial_\tau^2 V_P + \left( k^2 - \frac{\partial_\tau^2 \bar I}{\bar I} \right) V_P = 0 \; .
\label{EOM-V}
\end{equation}
The time dependence is often assumed to be $\bar I \propto a^n \propto (-\tau)^{-n}$ during inflation (quasi de Sitter). In this case, one has $\partial_\tau^2 \bar I / \bar I = n(n+1)/\tau^2$ and thus the solution of the mode function to \eqref{EOM-V} is
\begin{equation}
V_P (\tau , k) = i \, \frac{\sqrt{\pi}}{2} \sqrt{-\tau} \, H_{n + \frac{1}{2}}^{(1)} \left( - k \tau \right) 
\; ,
\label{Vlam-sol}
\end{equation}
where $H_\nu^{(1)}$ is the Hankel function of the first kind, and the arbitrary phase is fixed such that $V_{P}$ becomes real in the small $k$ limit with positive $n$.

By defining the magnetic field as $\vec{B} \equiv \frac{\bar I}{a^2} \vec{\nabla} \times \vec{A}$ and its power spectrum $P_B$ as usual
\begin{equation}
\left\langle \vec B (\tau ,\vec x) \cdot \vec B (\tau, \vec y) \right\rangle = \int \frac{dk}{k} \, \frac{\sin \left( k \vert \vec x - \vec y \vert \right)}{k \vert \vec x - \vec y \vert} \, P_B (\tau, k) \; ,
\end{equation}
and the same for the electric field $\vec{E} \equiv \frac{\bar I}{a^2} \partial_\tau \vec{A}$, 
we find
\begin{eqnarray}
P_B (\tau, k) 
& \simeq & \frac{2^{\vert 2 n + 1 \vert - 2} \, \Gamma^2\left( \big\vert n + \frac{1}{2} \big\vert \right)}{\pi^3} \, H^4 \left( \frac{k}{aH} \right)^{5 - \vert 2 n + 1 \vert}
\; ,
\label{PB-inf}\\
P_E (\tau, k)
& \simeq & 
\frac{2^{\vert 2 n - 1 \vert - 2} \, \Gamma^2\left( \big\vert n - \frac{1}{2} \big\vert \right)}{\pi^3} \, H^4 \left( \frac{k}{aH} \right)^{5 - \vert 2 n - 1 \vert}
\; ,
\label{PE-inf}
\end{eqnarray}
well outside the Hubble radius $k \ll aH$. While these expressions are valid only during inflation, the scale dependence is fully determined by \eqref{PB-inf} at all times. Therefore, a scale-invariant spectrum of magnetic fields can be realized for the cases $n= 2$ and $n=-3$, with amplitude given by
\begin{equation}
P_B \simeq \frac{9 H^4}{2\pi^2} \; , \qquad
n=2,-3 \; ,
\label{PB-inf-SI}
\end{equation}
where $H$ is evaluated during inflation. If $\bar I$ becomes unity to recover the standard EM at the end of inflation, and the magnetic field simply redshifts thereafter as $B^2 \propto a^{-4}$, then the resultant amplitude at the present time $t_{\rm now}$ becomes~%
\footnote{Since our primary aim is to account for the results of blazar observations by magnetic fields in the extragalactic regions, $t_{\rm now}$ can be taken as the time when electrons and positrons pass through such regions, corresponding to the redshift between $z = 0$ and $z \lesssim 0.5$ \cite{Neronov:1900zz, Tavecchio:2010mk, Dolag:2010ni, Essey:2010nd, Taylor:2011bn, Takahashi:2013lba, Finke:2013tyq, Chen:2014rsa}. This would only result in an extra factor $(1+z)^2 \sim 1 - 2.25$ in \eqref{PB-now} and would not change the order estimate.}
\begin{equation}
\begin{aligned}
\left( \frac{d B^2}{d\ln k} \right)^{1/2} (t_{\rm now}) = & P_B^{1/2} (t_{\rm now})
= \left( \frac{a_{\rm end}}{a_{\rm now}} \right)^{2} P_B^{1/2}(t_{\rm end})
\\ \simeq &
 8.5 \cdot 10^{-17} \, {\rm G}
\cdot \left( 5.1 \cdot 10^{23} \right)^{\frac{2w}{3(1+w)}} \\ & 
\times \left( \frac{g_*}{106.75} \right)^{\frac{2}{3(1+w)}} 
\left( \frac{H_{\rm inf}}{10^{12} \, {\rm GeV}} \right)^{\frac{2(1+3w)}{3(1+w)}}
\left( \frac{T_{\rm reh}}{10^9 \, {\rm GeV}} \right)^{\frac{2(1-3w)}{3(1+w)}}
\; ,
\end{aligned}
\label{PB-now}
\end{equation}
where $T_{\rm reh}$ is the reheating temperature, $g_*$ is the number of relativistic degrees of freedom at the time of reheating, and 
in deriving \eqref{PB-now} we assume that the universe goes through a period of inflaton oscillation with the (averaged) equation of state $w$ after inflation until the completion of reheating.

Eq.~\eqref{PB-now} appears to suggest that scale-invariant magnetic fields with the observed amplitude $B \gtrsim 10^{-15} \, {\rm G}$ can be obtained without much difficulty. However, as we briefly summarize in the following subsections, and as has already been known, there are observational constraints that must be respected, and they severely limit the present amplitude of magnetic fields of primordial origin.
Moreover we should emphasize that \eqref{PB-now} is valid only when the production of magnetic fields ends at the end of inflation and thus it does \emph{not} apply to cases with post-inflationary evolution, which we consider in Sec.~\ref{sec:increasing} below.

\subsection{Effective coupling to charged particles}
\label{subsec:eff-coupling}

Our prescription is to preserve the gauge invariance in the electromagnetic sector. This restricts the form of interactions to the standard model particles as ${\cal L}_{\rm EM} = e A_\mu J^\mu$, where $e$ is the electric charge and $J^\mu$ is the EM current, constructed by canonically normalized fields. In particular, $J^\mu$ cannot contain the kinetic function $I$ in \eqref{Lag-EM} due to the symmetry. Therefore once the EM field is canonically normalized as $A_\mu = A_\mu^{c} / \bar I$, we observe that the effective coupling is renormalized as
\begin{equation}
e_{\rm eff} \equiv \frac{e}{\bar I} \; .
\end{equation}
To assure perturbative calculations in the standard model sector, $e_{\rm eff} < 1$ is necessary throughout the evolution of the observable quantities.
On the other hand, in order to recover standard electromagnetism, we require $I \rightarrow 1$ by the time of, at the very least, big bang nucleosynthesis (BBN). Most straightforward attempts of magnetic field production are of inflationary origin and implement $I \to 1$ at the end of inflation. However, in such a case, $\bar I$ has to be a decreasing function at least for some e-folds toward the end of inflation to realize $e_{\rm eff} < 1$. Then severe constraints due to significant contributions of the produced {\it electric} fields to the CMB fluctuations apply, and production of large-scale {\it magnetic} fields, which is parametrically smaller than the electric field amplitude, must be critically suppressed; this suppression is so severe that even the production only during the last one e-fold of inflation could violate the CMB constraints \cite{Fujita:2013pgp,Fujita:2014sna}.

In this paper we consider an alternative scenario, where $\bar I$ is an increasing function of time during inflation, but ensuring $\bar I > 1$ for the duration concerning the Hubble exit of observable quantities, and it drops to unity due to the post-inflationary dynamics. As can be seen from \eqref{PE-inf}, the magnetic fields dominate over the electric during inflation in this case, and thus the CMB constraints described above do not apply. Therefore this scenario avoids the strong coupling problem while circumventing the issue of CMB observations. However, as we will see in Sec.~\ref{subsec:problem}, this model has its own suppression of magnetic fields, namely, when $\bar I$ evolves from a large value to unity after inflation, the magnetic-field amplitude $\vec{B}$ receives a suppression corresponding to this big drop, which makes the produced magnetic fields tens of orders of magnitude smaller than the lower bound \eqref{Beff} from blazar observations (see \eqref{Bnow_concrete} for a concrete number). We will elaborate on this issue in Sec.~\ref{subsec:problem}.

\subsection{Constraints}
\label{subsec:constraints}

A naive estimate of the present amplitude of magnetic fields in the models of \eqref{Lag-EM} with $\bar I \propto a^{2}$ or $a^{-3}$ is given in \eqref{PB-now}. However it also needs to pass several consistency and observational constraints, which sometimes severely limit the validity of \eqref{PB-now}. At the background level, the energy density of the produced electromagnetic fields must be subdominant to the total energy density in order not to disrupt the background dynamics as well as due to the fact that the production is supported by some other background quantity. This amounts to requiring
\begin{equation}
\frac{1}{2} \big\langle \vec{E}^2 + \vec{B}^2 \big\rangle \ll 3 M_{\rm Pl}^2 H^2 \; ,
\label{backreaction}
\end{equation}
where the bracket denotes the average. For the form $\bar I \propto a^n$, we have $\langle \vec{B}^2 \rangle \ll \langle \vec{E}^2 \rangle$ for $n < -2$, and $\langle \vec{E}^2 \rangle \ll \langle \vec{B}^2 \rangle$ for $n > 2$. Hence in the case of $n = -3$, while the magnetic field is scale-invariant \eqref{PB-inf-SI}, the electric spectrum is strongly red, and thus the contribution to its energy density is dominated by the IR. For $n < -2$, $\langle \vec{B}^2 \rangle \ll \langle \vec{E}^2 \rangle \sim H^4 \exp\left( \vert 2n+4 \vert N_{\rm prod} \right)$, where $N_{\rm prod}$ is the number of e-folds for the duration of the production during inflation.
This leads, from \eqref{backreaction}, to $H / M_{\rm Pl} \ll \exp\left( - \vert n+2 \vert N_{\rm prod} \right)$, and thus severely limits the possibility of large-scale magnetic fields in high-energy inflationary models, as has been known in the literature.

In some limited cases, we can relate $N_{\rm prod}$ to the peak scale, thus determining the coherence length of the produced magnetic fields. For simplicity, we here limit our discussion to the case where the production ends at the end of inflation. Then $N_{\rm prod}$ corresponds to the time duration for which $\bar I$ changes as $\bar I \propto a^n$ during inflation, and \eqref{PB-inf} applies to the modes that cross the Hubble radius during these $N_{\rm prod}$ e-folds ($\bar I = {\rm const.}$ outside this duration is assumed). In the case $-3 < n < 2$, the magnetic-field spectrum is blue, and so $N_{\rm prod}$ is irrelevant for the peak scale, which is given by the modes crossing the Hubble radius around the end of inflation.
For $n=-3,2$, the spectrum is scale-invariant, thus having no peak, and the modes that cross the Hubble radius $N_{\rm prod}$ e-folds before the end of inflation are the largest-scale modes produced.
If $n< -3$ or $2<n$, the spectrum is red, in which case $N_{\rm prod}$ is related to the length scale of the produced magnetic fields via the usual relation \cite{Liddle-lyth:2009}~\footnote{Here, $\rho_{\rm inf}$ and $\rho_{\rm ref}$ correspond to the energy densities at the end of inflation and at the end of reheating, respectively }
\be
N_{\rm prod} = 53 + \ln \frac{k_{\rm peak}^{-1}}{{\rm Mpc}} 
+ \ln \frac{\rho_{\rm inf}^{1/4}}{10^{16} \, {\rm GeV}}
- \frac{1 - 3w}{3(1+w)} \ln \frac{\rho_{\rm inf}^{1/4}}{\rho_{\rm reh}^{1/4}} \; ,
\label{Nprod-kpeak}
\ee
where $k_{\rm peak}$ is the comoving peak scale of the produced magnetic field, and we have assumed pure de Sitter during inflation and post-inflationary inflaton oscillation period with effective equation of state $w$.
The second term in the right-hand side of \eqref{Nprod-kpeak} implies that a longer duration of production is required for a larger coherence length, while the third term means that $N_{\rm prod}$ can be smaller for a given peak scale if the inflationary energy scale is lowered. The last term in \eqref{Nprod-kpeak} comes from our ignorance of the reheating process. Let us note that this term vanishes if the universe behaves as radiation-dominated during inflaton oscillation ($w = 1/3$), as in the example in Sec.~\ref{subsec:jumpmodel}.
The scale $k_{\rm peak}=1$ Mpc$^{-1}$ corresponds to a mode with $N_{\rm prod} = N_{\rm CMB} + \log (k_{\rm CMB} / {\rm Mpc}^{-1}) \simeq N_{\rm CMB} - 3$,%
\footnote{Here, $N_{\rm CMB}$ is the number of e-foldings before the end of inflation when the modes at $k_{\rm CMB}$ exit the Hubble radius.}
where in the second equality we have chosen the CMB pivot scale $k_{\rm CMB}=0.05 \, \mbox{Mpc}^{-1}$ .

On the other hand, in the case of $n=2$, the hierarchy goes as $\langle \vec{E}^2 \rangle \ll \langle \vec{B}^2 \rangle \sim H^4 N_{\rm prod}$, translating \eqref{backreaction} to $H / M_{\rm Pl} \ll N_{\rm prod}^{-1/2}$, which is easily satisfied as $H \lesssim {\cal O}(10^{-5}) M_{\rm Pl}$ for most, if not all, inflationary models.
In fact the constraint \eqref{backreaction} can be met easily for $n>0$; the difficulty comes in to avoid the strong coupling problem discussed in Sec.~\ref{subsec:eff-coupling}. To preserve the weak coupling of photon to the charged particles, $e / \bar{I}$ needs to be smaller than unity at all times. Therefore, for $n > 0$, we require $\bar{I} \gtrsim 1$ at the end of inflation. As discussed in Sec.~\ref{subsec:problem}, the transition from $\bar{I} > 1$ to $\bar{I} \to 1$, to recover the standard electromagnetism, inevitably leads to an unobservable amplitude of the magnetic fields.

Another important constraint is the consistency with the CMB observations. The curvature perturbation $\zeta$ evolves outside the Hubble radius due to the non-adiabatic pressure provided by the produced electromagnetic field. During inflation, this sourced contribution $\zeta_{\rm EM}$ can be found as \cite{Fujita:2013pgp,Fujita:2014sna}
\begin{equation}
\zeta_{\rm EM} \left( t, \vec{x} \right) = - \frac{2H}{3 \left( \rho + p \right)} \int^t dt' \, \left[ \vec{E}^2 \left( t', \vec{x} \right) + \vec{B}^2 \left( t', \vec{x} \right) \right] \; ,
\end{equation}
which is derived from the super-Hubble evolution $\dot\zeta = - H \delta p_{\rm nad} / (\rho + p)$, where the non-adiabatic pressure $\delta p_{\rm nad}$ is now sourced by the electromagnetic fields.
In the case of $n<-2$, since the electromagnetic energy is dominated by the electric field, the observational constraints coming from the curvature power spectrum and local-type non-Gaussianity functions place bounds on the electric power spectrum \cite{Fujita:2013pgp}
\begin{equation}
P_E \lesssim H^4 \,
\frac{3 \left( \vert n \vert - 2 \right)}{2^2 \pi^2} \, \frac{P_\zeta}{P_0} \times
\min \left[ \left( \frac{1}{P_\zeta D_n} \right)^{1/2} , 
\left( \frac{\Delta f_{\rm NL, local}^{\rm obs}}{P_\zeta D_n} \right)^{1/3} 
\right] \; , \qquad
n < -2
\label{CMBconst-Idec}
\end{equation}
where $P_\zeta \simeq 2.2 \times 10^{-9}$ \cite{Ade:2015lrj}, $\Delta f_{\rm NL, local}^{\rm obs} \sim {\cal O}(5)$ \cite{Ade:2015ava}, $P_0 \equiv H^2 / (8\pi^2 \epsilon M_{\rm Pl}^2)$, and $D_n \equiv \left({\rm e}^{(2 \vert n \vert - 4) (N_{\rm prod} - N_{\rm CMB})} - 1 \right) / ( 2 \vert n \vert - 4)$ with $N_{\rm CMB}$ the number of e-folds after the CMB modes exit the Hubble radius until the end of inflation. 
The first term in $\min[\dots]$ in \eqref{CMBconst-Idec} comes from the requirement that $P_\zeta$ induced by the produced EM fields should not exceed the observed value of $P_\zeta$, and the second term is that the level of the induced scalar non-Gaussianity should be below the observational bound. We have rewritten these conditions as an upper bound for $P_E$ in \eqref{CMBconst-Idec}.
Since $P_B \sim {\rm e}^{- 2 N_{\rm CMB}} P_E$, the amplitude of the magnetic field is strongly suppressed by the exponential factor ${\rm e}^{- 2 N_{\rm CMB}}$ in the case of $n < -2$. In particular, in the case of scale-invariant magnetic field $n = -3$, and assuming $N_{\rm prod} - N_{\rm CMB} \geq 0$, \eqref{CMBconst-Idec} translates to
\begin{eqnarray}
P_B & \lesssim & 
3 \cdot 10^{-3} H^4 \, {\rm e}^{- 2 N_{\rm CMB}} \, \frac{P_\zeta}{P_0} \times
\min \left[ 3 \cdot 10^{4} \, {\rm e}^{-(N_{\rm prod} - N_{\rm CMB})} ,  \,
2 \cdot 10^3 \, {\rm e}^{- \frac{2}{3} (N_{\rm prod} - N_{\rm CMB})}
\right] 
\nonumber\\
& \lesssim &
90 H^4 \, {\rm e}^{-(N_{\rm prod} + N_{\rm CMB})} \; , \qquad
n = -3 \; ,
\label{PB-CMB-Idec}
\end{eqnarray}
where in the last inequality we have taken, for the purpose of demonstration, $N_{\rm prod} - N_{\rm CMB} \gtrsim 9$, and $P_0 \simeq P_\zeta$, which is the case when the observed curvature perturbation is dominated by the vacuum fluctuations of inflaton. This constraint suppresses the final amplitude of the magnetic field by ${\rm e}^{-(N_{\rm prod} + N_{\rm CMB})/2}$ as compared to \eqref{PB-now}, making it completely unobservable.
The expression \eqref{PB-CMB-Idec} implicitly assumes that the production of the electromagnetic fields has started at the time when the CMB modes cross the Hubble radius. Even without this assumption, however, it has been shown in  a model-independent manner \cite{Fujita:2014sna} that inflationary generation of large-scale magnetic fields is extremely difficult even if the production is limited only to a few last e-folds toward the end of inflation, as long as $n < 0$.

The situation is significantly different for positive values of $n$. For $n \ge 2$, the energy in the electromagnetic fields is dominated by the magnetic one, instead of the electric one, and thus the CMB constraint applies directly to $P_B$. For the values $n > 2$, due to the duality $I \leftrightarrow 1/I$ with $\vec{B} \leftrightarrow - \vec{E}$ and $\vec{E} \leftrightarrow \vec{B}$ \cite{Giovannini:2009xa}, the CMB constraint takes exactly the same form as \eqref{CMBconst-Idec} but with $P_E$ replaced by $P_B$. In this case, however, the magnetic field spectrum becomes strongly red, and hence the phase-space contributions from the IR quickly violate the bound. In the case of $n=2$, which gives a scale-invariant magnetic spectrum, the CMB constraint reads \cite{Barnaby:2012tk}
\begin{eqnarray}
N_{\rm prod} - N_{\rm CMB} & \lesssim &
\min \left[ \frac{5 \cdot 10^{-3} P_\zeta}{\left( P_0 N_{\rm CMB} \right)^2} , \, 
\frac{8 \cdot 10^{-4} P_\zeta^2}{\left( P_0 N_{\rm CMB} \right)^3} \, \Delta f_{\rm NL, local}^{\rm obs}
\right] 
\nonumber\\
& \lesssim &
\frac{2 \cdot 10^{6}}{N_{\rm CMB}^3} \; , \qquad
n = 2 \; ,
\label{CMBconst-Iinc}
\end{eqnarray}
where in the last inequality the standard case of the curvature perturbation, $P_0 \simeq P_\zeta$, is assumed for illustrative purposes.
The first line of \eqref{CMBconst-Iinc} comes, similarly to \eqref{CMBconst-Idec}, from the requirement that the EM-induced curvature perturbations are smaller than the observed value of $P_\zeta$ (the first term in $\min[\dots]$) and below the bound on non-Gaussianity (the second term).
For a given value of $N_{\rm CMB} \sim 50 - 60$, \eqref{CMBconst-Iinc} places an upper bound on the duration $N_{\rm prod}$ of the production of the electromagnetic fields. Therefore, if this were the end of the story, this would not immediately constrain the strength of the magnetic fields, and thus the resultant amplitude \eqref{PB-now} would appear to hold. However, we now need to take into consideration the effective electromagnetic coupling, discussed in Sec.~\ref{subsec:eff-coupling}. Although the normalization of $\bar I$ does not affect the constraint \eqref{CMBconst-Iinc}, we must assume $\bar I \gg 1$ at the end of inflation for the weak coupling since $\bar I$ increases during inflation for $n>0$. To recover the standard electromagnetism, it is mandatory to implement the mechanism to realize $\bar I \rightarrow 1$ after inflation. As we will show in Sec.~\ref{subsec:problem}, this transition severely suppresses the resulting amplitude of scale-invariant part of the magnetic spectrum in this scenario, making it completely unobservable also in the case $n > 0$.

We should note that all the CMB constraints discussed above, \eqref{CMBconst-Idec}, \eqref{PB-CMB-Idec} and \eqref{CMBconst-Iinc}, are valid only when $N_{\rm prod} > N_{\rm CMB}$. In the opposite case, namely when the production of the electromagnetic fields has not occurred at the time of the CMB modes crossing the Hubble radius, as considered in \cite{Ferreira:2013sqa,Ferreira:2014hma,Kobayashi:2014sga,Fujita:2016qab}, these bounds do not immediately apply. However, as shown in \cite{Fujita:2014sna}, if the production ends at the end of inflation and if $\bar I$ is decreasing at least for a few e-folds toward the end of inflation, then the realization of large-scale magnetic fields is extremely difficult unless one constructs a  dedicated model of very low inflationary scale.
According to our analysis above, on the other hand, the opposite case of an increasing $\bar I$ can evade the CMB constraints without much difficulty. Our results in the following sections are to close this window in the context of large-scale magnetic fields that could explain the blazar observations.

\section{Increasing $I$ without a strong coupling problem}
\label{sec:increasing}

The brief summary of constraints in Sec.~\ref{subsec:constraints} demonstrates that the production of magnetic fields is strongly suppressed in the case of decreasing $\bar I$ during inflation. This is due to the substantial growth of {\it electric} field amplitudes. Since electric fields are produced with larger amplitudes than those of magnetic fields, a  limit on the former restricts the latter even more severely. 

\begin{figure}
\centering
\includegraphics[scale=0.6]{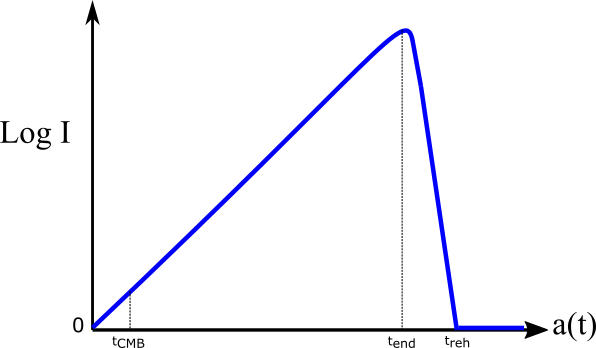}
\caption{Schematic Evolution of $I$ }
\label{fig:schematic}
\end{figure}

One obvious way to circumvent this is to have increasing $\bar I$ during the production of magnetic fields. In this case, magnetic fields are amplified more than electric ones, and therefore the severe constraints discussed in Sec.~\ref{subsec:constraints} effectively become irrelevant.
At the same time, however, we need to ensure weak couplings to the standard model particles, as in Sec.~\ref{subsec:eff-coupling}, throughout the period during which any observable quantities are generated. The earliest time for this concern is the time when the modes relevant for CMB observations exit the Hubble radius, denoted $t_{\rm CMB}$.
In order to ensure weak coupling after the CMB observables become classical, we require
\begin{equation}
\label{Icond1}
\bar I > 1 \; , \qquad t > t_{\rm CMB} \; ,
\end{equation}
where $t_{\rm CMB}$ is the time when the CMB modes exit the Hubble radius during inflation, and we are not concerned with the behavior of $\bar I$ before $t_{\rm CMB}$.
In this section, we are interested in an increasing $\bar I$ during inflation that leads to a scale-invariant magnetic fields and thus require
\begin{equation}
\label{Icond2}
\bar I \propto a^2 \; , \quad t_{\rm CMB} < t < t_{\rm end} \; ,
\end{equation}
denoting the time at the end of inflation by $t_{\rm end}$.
After inflation, the value of $\bar I$ must recover the standard normalization, i.e.
\begin{equation}
\label{Icond3}
\bar I \rightarrow 1 \; , \quad t \to t_{\rm reh} \; ,
\end{equation}
where we conservatively assume that this transition occurs by the time at the completion of reheating $t_{\rm reh}$. Our conclusion would not change by pushing this time to that of BBN. A qualitative sketch of the evolution of $\bar I$ that satisfies  \eqref{Icond1},\eqref{Icond2}, \eqref{Icond3}, is given in Figure \ref{fig:schematic}. 

With this setup, the power spectrum of the produced magnetic fields at the end of inflation is given by \eqref{PB-inf}, namely
\begin{equation}
P_B(t=t_{\rm end}) \simeq \frac{9 H_{\rm inf}^4}{2\pi^2} \; .
\label{PBinf-Iinc}
\end{equation}
The important difference between this scenario and the well-studied case with decreasing $\bar I \propto a^{-3}$, which also gives a scale-invariant spectrum of magnetic fields, is that the energy density of the gauge field is now dominated by the magnetic fields, $\rho_B \gg \rho_E$. This implies that the constraints discussed in Sec.~\ref{subsec:constraints} are imposed directly on the magnetic fields,
with the electric fields negligible.
The bounds on their energy density, \eqref{backreaction}, now reads~%
\footnote{Strictly speaking, $\rho_B$ must not exceed the kinetic energy of the producing fields, not necessarily inflaton. Since we so far have not specified the dynamics of the latter, we here impose a more conservative condition, \eqref{backreac-Iinc}.}
\begin{equation}
\frac{\rho_B}{3M_{\rm Pl}^2 H_{\rm inf}^2} \simeq \frac{3H_{\rm inf}^2}{4\pi^2 M_{\rm Pl}^2} \, N_{\rm prod} \ll 1 \; ,
\label{backreac-Iinc}
\end{equation}
at the end of inflation, where $N_{\rm prod}$ is the total number of e-folds from the time when the largest-scale modes of the gauge field exited the Hubble radius to that of the end of inflation.
On the other hand, 
if the magnetic field production is already effective at the time when the CMB modes exit the Hubble radius, the curvature perturbation induced by the produced magnetic fields must be subdominant to the standard vacuum fluctuations, $P_\zeta^{\rm sourced} < P_\zeta \cong 2.2 \cdot 10^{-9}$, leading to, \eqref{CMBconst-Iinc},
\begin{equation}
N_{\rm prod} - N_{\rm CMB}
\lesssim
\frac{2 \cdot 10^{6}}{N_{\rm CMB}^3} \; ,
\label{curvpert-Iinc}
\end{equation}
where $N_{\rm CMB}$ is the number of e-folds from the time of the CMB modes' Hubble radius crossing to the end of inflation.
Notice that generically for $H_{\rm inf} \lesssim 10^{-5} M_{\rm Pl}$, the condition \eqref{curvpert-Iinc} is more stringent than \eqref{backreac-Iinc}.

At a superficial level, the requirement \eqref{curvpert-Iinc} is not difficult to meet.%
\footnote{Moreover, if we assume that the dominant part of $P_\zeta$ comes from other dynamics such as curvaton, then the condition \eqref{curvpert-Iinc} becomes even more loose. Also, while inclusion of higher-order correlation functions would make the bound on $N_{\rm prod} - N_{\rm CMB}$ somewhat stronger as shown in \cite{Fujita:2013pgp}, there would still be a viable parameter region.}
Hence, if the magnetic fields were only to be redshifted after inflation from the value given in \eqref{PBinf-Iinc}, then one could have a sufficiently large amplitude of magnetic fields to explain the blazar observations. However, we remind ourselves that, under the current scenario with increasing $\bar I$ during inflation, it is inevitable that $\bar I \gg 1$ at the end of inflation in order to avoid the strong coupling problem. A subsequent evolution needs to realize $\bar I \rightarrow 1$ for the normalization consistent with the present electromagnetism. 
Next in Sec.~\ref{subsec:problem} we show that this transition, quite generally, leads to a large additional suppression of the scale-invariant magnetic fields, invalidating this mechanism for the generation of large-scale magnetic fields. Then in Sec.~\ref{subsec:jumpmodel} we study a concrete realization of the scenario discussed here.

\section{Fatal problem from post-inflationary transition}
\label{subsec:problem}

Now that we have discussed the mechanism to avoid the strong-coupling problem in the previous section, we come to computing the final amplitude of the produced magnetic fields.
Given the inflationary solution \eqref{Vlam-sol} for the gauge field, it is more convenient  and intuitive to switch to the (modified) Maxwell equations for the subsequent evolution. 
The equation of motion for the gauge potential, \eqref{EOM-V}, is equivalent to the set of equations
\begin{equation}
\begin{aligned}
\partial_\tau \left( \frac{\vec B^c}{\bar I} \right) \, = - \vec{\nabla} \times \left( \frac{\vec E^c}{\bar I} \right) \; , \quad
\partial_\tau \left( \bar I E^c \right) = \vec{\nabla} \times \left( \bar I \vec B^c \right) \; ,
\end{aligned}
\label{Maxwell}
\end{equation}
where the ``comoving'' magnetic and electric fields are defined as (in the Coulomb gauge $A_0 = \vec{\nabla} \times \vec A = 0$)
\begin{equation}
\vec B^c \equiv \bar I \, \vec{\nabla} \times \vec{A} \; , \quad
\vec E^c \equiv - \bar I \, \partial_\tau \vec{A} \; .
\label{BcEc}
\end{equation}
Note that $\vec{B}^c$ and $\vec{E}^c$ are related to the ``physical'' fields $\vec{B}$ and $\vec{E}$ as $\vec{B}^c  = a^2 \vec{B}$ and $\vec{E}^c = a^2 \vec{E}$. The known duality $\bar I \leftrightarrow 1 / \bar I$ with $\vec{B}^c \rightarrow - \vec{E}^c$ and $\vec{E}^c \rightarrow \vec{B}^c$ is manifest in \eqref{Maxwell}.

Decomposing $B^c$ and $E^c$ into the circular polarization states along $e_i^P$ as in \eqref{A-FT}, the Maxwell equations \eqref{Maxwell} are rewritten in the Fourier space as
\begin{equation}
\partial_\tau \left( \frac{B_P^c}{\bar I} \right) = - P k \, \frac{E_P^c}{\bar I} \; , \quad
\partial_\tau \left( \bar I E_P^c \right) = P k \bar I B_P^c \; ,
\label{Maxwell2}
\end{equation}
where $P=\pm$ is the photon polarization. The inflationary solution \eqref{Vlam-sol} is translated to
\begin{equation}
\begin{aligned}
B_P^c 
= i P k \, \frac{\sqrt{\pi}}{2} \sqrt{-\tau} \, H_{n + \frac{1}{2}}^{(1)} \left( - k \tau \right) \; , \quad 
E_P^c 
= - i k \, \frac{\sqrt{\pi}}{2} \sqrt{-\tau} \, H^{(1)}_{n - \frac{1}{2}} (-k\tau) \; .
\end{aligned}
\label{BcEc-infsol}
\end{equation}
Since $B_P^c \gg E_P^c$ for $n > 1/2$ ($\vert E_P^c / B_P^c \vert_{\tau=\tau_{\rm end}} \sim k / (a_{\rm end} H_I) \ll 1$ at the end of inflation), the first equation in \eqref{Maxwell2} can immediately be solved as
\begin{equation}
B_P^c \propto \bar I \; ,
\label{Bsol-Idep}
\end{equation}
with the proportionality constant given by the initial condition. Let us emphasize that \eqref{Bsol-Idep} is valid whenever $B_P^c \gg E_P^c$ holds, namely all the time after a mode of interest exits the Hubble radius during inflation.
Moreover, the other part of the solution of \eqref{Maxwell2} for $B_P^c$, which comes from the electric field contribution, is more blue-tilted than the part in \eqref{Bsol-Idep} by additional $k^2$, and therefore it cannot contribute to scale-invariant magnetic spectra in the models of increasing $I$.

The solution \eqref{Bsol-Idep} immediately poses a question concerning the final amplitude of scale-invariant magnetic fields. Regardless of the detailed dynamics of the magnetic fields, we require $\bar I \rightarrow 1$ after some time in order to recover standard electromagnetism. When taken in conjunction with \eqref{Bsol-Idep}, which is valid already during inflation, this then implies that  $\vec B^c (t_{\rm now} , k) = \vec B^{c} (t_*, k) / \bar I_*$, where $*$ denotes the time of Hubble radius crossing of given modes, or in terms of the physical magnetic field,%
\footnote{This expression is valid for the modes on scales large enough to be unaffected by the magnetohydrodynamics during the radiation-dominated period.}
\begin{equation}
\vec B (t_{\rm now}, k) = \frac{1}{\bar I_*} \left( \frac{a_*}{a_{\rm now}} \right)^2 \vec B (t_*, k) \; .
\end{equation}
This expression encodes an unavoidable suppression of the scale-invariant part of physical magnetic fields. For example, consider ``now" to be immediately after the end of inflation, and $t_*$ to be moment before the end of inflation. In this case, $ \frac{a_*}{a_{\rm now}} \sim 1$ but $\vec B (t_{\rm now}, k)$ is suppressed by the large factor ${\bar I_*} \gg 1$ (by virtue of $I$ growing as $a^2$ during inflation). Alternatively, for $t_*$ set to be 60 e-folds before the end of inflation, and leaving ``now" to be immediately after inflation, we have ${\bar I_*} \sim 1$ but $\vec B (t_{\rm now}, k)$ is suppressed by a factor of $ \left( \frac{a_{\rm now}}{a_*} \right)^2 \sim e^{120} \sim 10^{52}$.

We can make this more precise by by evaluating the spectrum of magnetic fields today. Using \eqref{PBinf-Iinc}, one arrives at
\begin{equation}
\begin{aligned}
\frac{d \big\langle \vec{B}^{\,2} \big\rangle}{d \ln k} (t_{\rm now}) & = \left[ \frac{1}{\bar I_*} \left( \frac{a_*}{a_{\rm now}} \right)^2 \frac{3 H_I^2}{2 \pi} \right]^2
\approx
\left[ 1.2 \cdot 10^{-13} \, {\rm G} \, \frac{{\rm e}^{-2N_*}}{\bar I_*}
\, \frac{H_I}{10^{12} \, {\rm GeV}} \right]^2 \; ,
\end{aligned}
\label{Bspecnow-Iinc}
\end{equation}
for scale-invariant magnetic fields, where $N_*$ is the number of e-folds from the Hubble radius crossing to the end of inflation, $t_{\rm now}$ denotes the present time, and
we have assumed that the total energy density effectively goes as $\propto a^{-4}$ between the end of inflation and the completion of reheating (i.e.~during inflaton oscillation).%
\footnote{This is marginally the most optimistic scenario, assuming that the effective potential of the inflaton after inflation is $\propto \phi^4$ or lower-order. If it is $\phi^p$ with $p<4$, then the background energy density decreases slower than $a^{-4}$, the magnetic field energy density compared to the background decreases faster, and therefore the present magnetic field amplitude becomes even smaller than \eqref{Bspecnow-Iinc}.}

If one assumes $\bar I$ evolves as $\propto a^2$ throughout inflation, the optimal scenario is $N_* = N_{\rm CMB} \approx 50 - 60$ with $\bar I_* = 1$, which amounts to $\vert \vec{B} (t_{\rm now}) \vert \approx ( 10^{-65} - 10^{-57} ) \, {\rm G} \times ( H_I / 10^{12} {\rm GeV})$. On the other hand, even if we seek only to explain the blazar observations and assume that $\bar I$ starts behaving as $a^2$ when the modes corresponding to $\sim 1 \, {\rm Mpc}$ cross the Hubble radius, i.e.~$N_* = N_{1 {\rm Mpc}} \approx N_{\rm CMB} - 3$ with $\bar I_* = 1$, we obtain $\vert \vec{B} (t_{\rm now}) \vert \approx ( 10^{-62} - 10^{-54} ) \, {\rm G} \times ( H_I / 10^{12} {\rm GeV})$, which is substantially smaller than the values necessary for the blazar observations.%
\footnote{Note that in the cases of $N_* < N_{\rm CMB}$, the constraints discussed in Sec.~\ref{subsec:constraints} do not directly apply. In fact, these cases would make it easier to satisfy the CMB constraints, as the production has not started at the time when the CMB modes cross the Hubble radius. Nonetheless, as shown here, no sufficiently large amplitude of magnetic fields can be realized in such scenarios.}

We should emphasize that the discussion above is quite general and does not rely on any specific model to realize the jump of $\bar I$ after inflation. The only underlying assumptions are (i) scale-invariant magnetic fields at the scales of comoving $\sim 1 \, {\rm Mpc}$, (ii) in the regime $\vert \vec B \vert \gg \vert \vec E \vert$ on superhorizon scales to avoid strong backreactions, and (iii) weak coupling of the photon to the Standard Model. Therefore, this result, combined with the existing knowledge of the difficulty in the case of decreasing $I$, severely narrows down the range of viable scenarios of the $I^2 F^2$ model for large-scale magnetogenesis of primordial origin.

\section{A concrete model}
\label{subsec:jumpmodel}

For the sake of completeness we now demonstrate a concrete scenario to realize the mechanism of Sec.~\ref{sec:increasing} in which the function $\bar I$ increases during inflation with $\bar I \gg 1$ at the end of inflation and then transits back to unity some time later. To this end, we promote $\bar I(t)$ to a function of both the inflaton $\varphi$ and a spectator field $\chi$ as
\begin{equation}
I(\varphi , \chi) = \exp \left( - \alpha \, \frac{\varphi^2}{M_{\rm Pl}^2} + \frac{\chi}{M} \right) \; ,
\label{form-I}
\end{equation}
where $M$ is a parameter with units of mass. The positive dimensionless parameter $\alpha$ is chosen to lead to an appropriate time dependence of $\bar I \equiv \sqrt{\langle I^2 \rangle}$ during inflation; in order to achieve $\bar I \propto a^n$ with the inflaton potential $V(\varphi) \propto \varphi^p$, it takes $\alpha = n / (2p)$.  The scenario we implement is to stabilize $\chi$ at $\chi = \chi_0$ during inflation, with $\chi_0 > M$,  and then after inflation $\chi$ moves to $\chi = 0$. This then realizes the desired transition of $\bar I$ 
\begin{equation}
\bar I (t_{\rm end}) \simeq \exp \left( \frac{\chi_0}{M} \right) \gg 1 \; , \quad
\bar I (t \rightarrow t_{\rm reh}) \rightarrow 1 \; .
\end{equation}

This can be realized in a variety of ways. One option is to make use of a `landscape' of vacua of $\chi$, with a transition to a new vacuum occurring due to non-trivial dynamics at the end of inflation or during preheating. The transition could occur via quantum effects, either by a tunneling to the true vacuum or an unstable growth of quantum fluctuations of $\chi$. The transition could instead occur as a ``classical tunneling'' event, following the unstable growth of \emph{classical} fluctuations (e.g. during preheating), similar in spirit to the ``geometric destabilization'' of moduli proposed in \cite{Renaux-Petel:2015mga}. In both the quantum and classical ``tunneling'' cases, the field $\chi$ becomes highly stochastic and bubbles of true vacuum will spontaneously nucleate. The idea of destabilization of moduli fields has been studied extensively in the literature, specifically in the so-called hybrid inflation models \cite{Linde:1993cn, Linde:1991km, Copeland:1994vg, Lazarides:2000ck}.

An alternative mechanism, which we develop in this section, is to consider a shift in the vacuum of $\chi$ due to the change in the vacuum expectation value (VEV) of $\varphi$, similar in spirit to the ``moduli vacuum misalignment'' phenomenon developed in \cite{Cicoli:2016olq}. This requires no recourse to a landscape of vacua, and only requires that the potential of $\chi$ exhibits at least a single minimum. The resulting dynamics is a global change in the VEV of $\chi$, avoiding complications due to bubbles and bubble wall collisions (as plague tunnelling). To this end, we introduce a coupling between $\chi$ and $\varphi$
\begin{equation}
{\cal L}_{\rm int} [\varphi, \chi] = - \frac{g^2}{2} \, \varphi^2 \left( \chi - \chi_0 \right)^2 \; .
\label{interaction}
\end{equation}
Therefore, the complete action of our model is
\begin{equation}
S = \int d^4x \sqrt{-g} \left[ \frac{M_{\rm Pl}^2}{2} R 
- \frac{1}{2} \left( \partial \varphi \right)^2 - V(\varphi) 
- \frac{1}{2} \left( \partial \chi \right)^2 - U(\chi)
+ {\cal L}_{\rm int} [\varphi , \chi]
- \frac{I^2(\varphi, \chi)}{4} F^2 \right]
\label{action}
\end{equation}
where $R$ is the Ricci scalar.

The exact form of $U(\chi)$ is not important for our general arguments, and in particular, the global structure of $U(\chi)$ would not play a role. In order for the mechanism to work, the bare mass of $\chi$ must be smaller than mass due to $g^2 \varphi^2$ at the end of inflation, i.e.
\begin{equation}
U_{\chi\chi} (\bar\chi_{\rm end}) \ll g^2 \varphi^2 \; ,
\end{equation}
and therefore $\chi$  must be stabilized by the interaction during inflation.  In order to have a jump from $\chi_0$ to a different value after inflation, i.e.~when $\varphi \rightarrow 0$, $U(\chi)$ must have a nearby minimum located at a value away from the interaction-stabilized value $\chi_0$, and for concreteness, we take this to be at the origin of $U(\chi)$.   These conditions are summarized as
\begin{equation}
U_{\chi\chi} (\chi_0) \ll g^2 \varphi^2 \; , \quad
U_{\chi} (0) = 0 \; , \quad U_{\chi\chi} (0) > 0 \; , \quad
U(0) \le U(\chi) \;\;\; {\rm for \; any \; } |\chi| \lesssim |\chi_0| \; .
\label{cond-U}
\end{equation}
The condition on the farthest-to-the-right implies that $U(0)$ should be a nearby local minimum. 
The potential could have other minima (or the global minimum) far from $\chi_0$, but this is not relevant to the perturbative physics we consider here. Moreover, self-interactions of $\chi$ are unimportant  for the background dynamics provided that they do not violate the conditions above. The time evolution of $I$ is schematically plotted in Figure \ref{fig:schematic}.

As an example, consider $\lambda \varphi^4$ inflation with a mass term for $\chi$:
\be
V(\varphi) = \frac{1}{4} \lambda \varphi^4 \;\ , \;\; U(\chi) = \frac{1}{2} m_\chi ^2 \chi^2 .
\ee
The effective potential for $\chi$ is given by
\be
U_{\rm eff} (\chi) = \frac{1}{2} m_\chi ^2 \chi^2 + \frac{g^2}{2} \varphi^2 (\chi- \chi_0)^2 . 
\ee
The effective potential is minimized at
\be
\chi_{*} = \frac{ g^2 \varphi^2}{m_{\chi}^2 +  g^2 \varphi^2} \, \chi_0 .
\ee
If $\chi$ is initially at $\chi=\chi_*$, then slow-roll inflation proceeds as usual, with the time-evolution of $\varphi$ given by $\varphi(N) \approx 2\sqrt{2} \, M_{\rm Pl} \sqrt{N+1}$, where $N$ is the number of e-folds before the end of inflation. Thus the value $\chi_*$ evolves as
\be
\chi_* (N) \simeq \chi_0 \left( \frac{1}{1 + 
\frac{m_{\chi}^2}{8 \, g^2 M_{\rm Pl}^2} \frac{1}{1+N}}  \right)  \; .
\ee
We immediately see that $\chi_* \approx \chi_0$ is constant during inflation,
as long as $m_\chi / g < M_{\rm Pl}$, and it is not until $\varphi$ decreases below $m_{\chi}/g$ that there is any significant time evolution of $\chi$. After inflation, both $\varphi$ and $\chi$ undergo a short period of oscillation, and $\varphi$ decays into standard model particles. Both fields quickly settle down to $0$.  

The form of $I$ required to have $\bar{I} \propto a^2$ during inflation is given by \eqref{form-I} with $\alpha = 1/4$,
\begin{equation}
I(\varphi , \chi) = \exp \left( - \frac{\varphi^2}{4 M_{\rm Pl}^2} + \frac{\chi}{M} \right) \; ,
\end{equation}
which evolves during inflation as,
\begin{equation}
\bar{I}(N) \simeq \exp \left[ - 2 (N+1) + \frac{\chi_0}{M} \right] \; , \qquad N > 0 \; .
\end{equation}
From this one can readily compute the spectrum of magnetic fields at late times.

\begin{figure}[h!]
\centering
\includegraphics[width=0.48\textwidth]{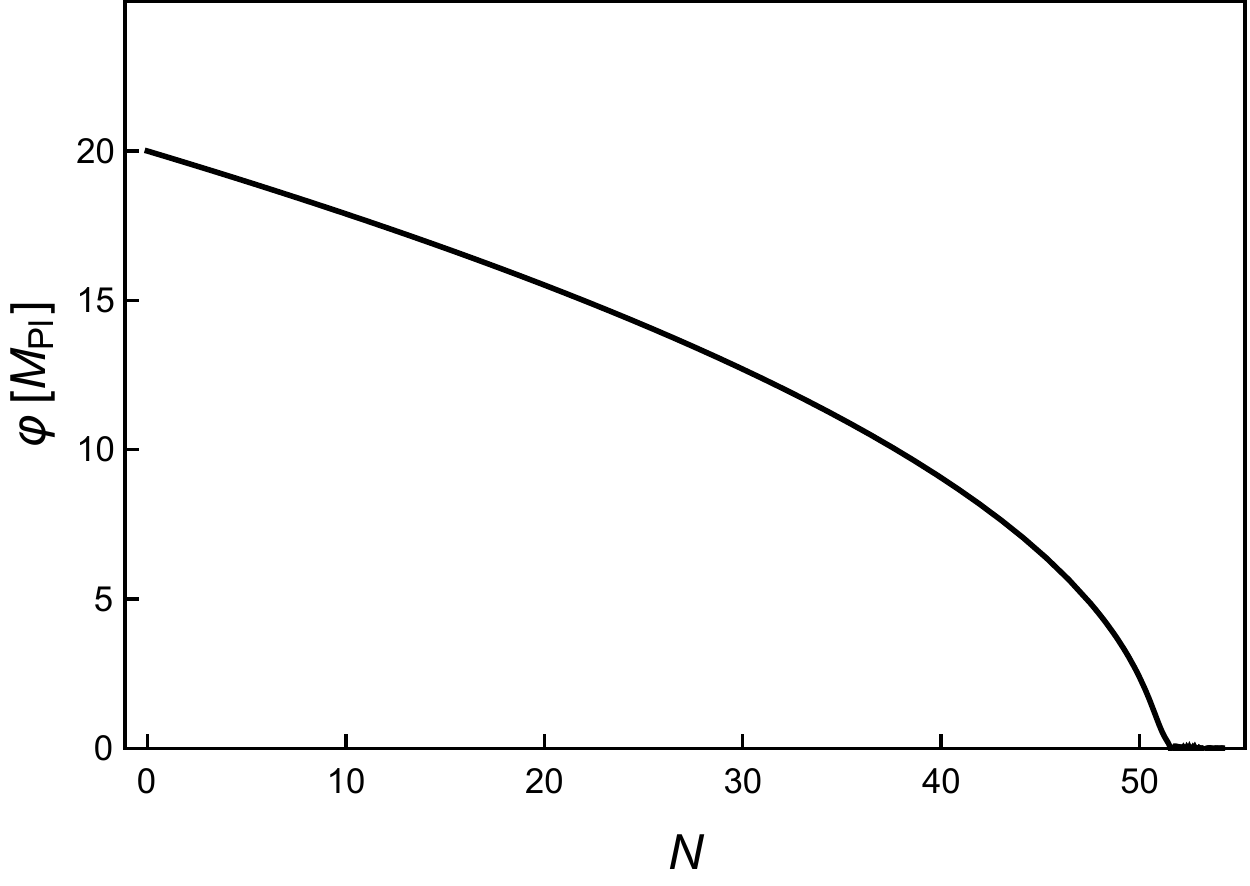} \hfill
\includegraphics[width=0.48\textwidth]{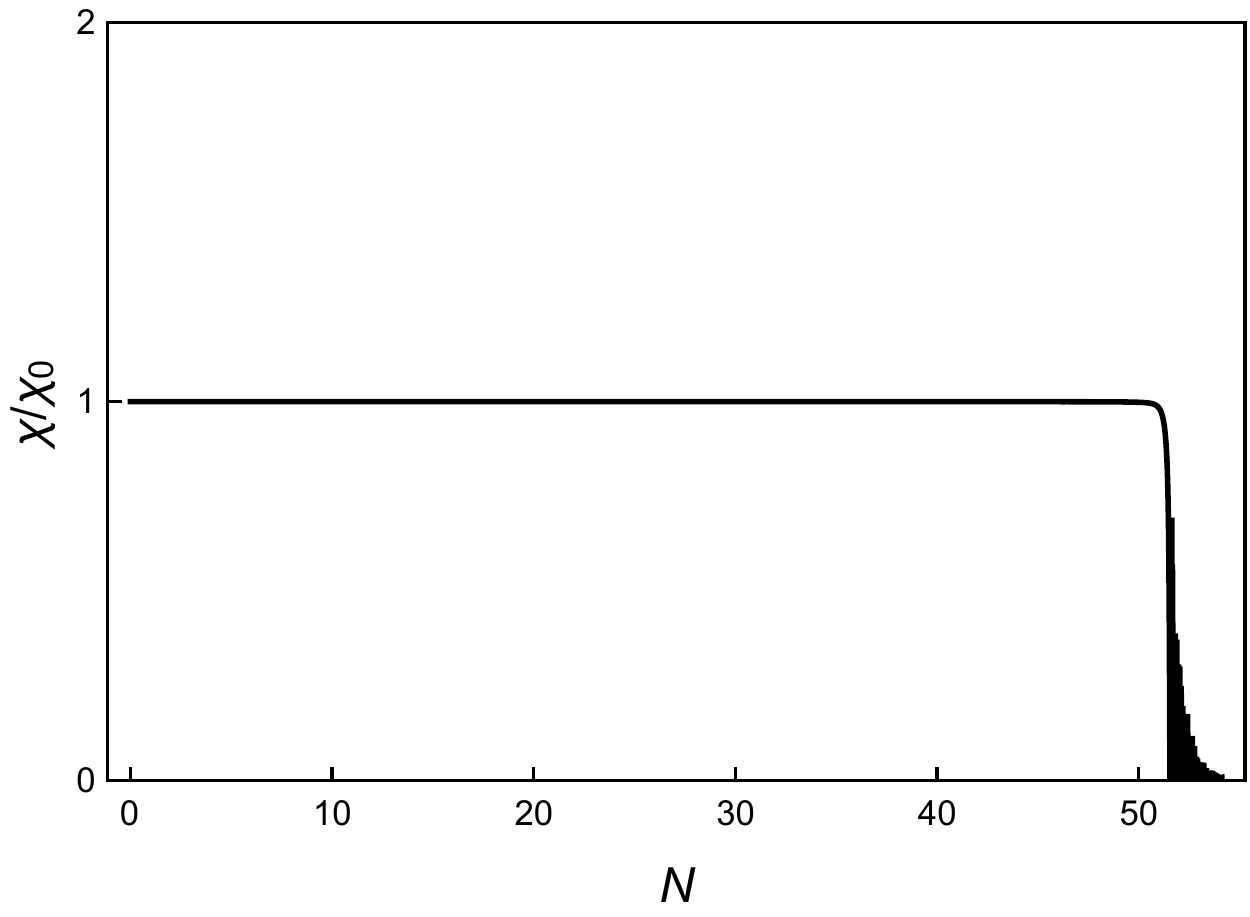} 
\caption{
{\bf [Left panel]} Evolution of $\varphi$, in Planck units. {\bf [Right panel]} Evolution of $\chi$, rescaled by $\chi_0$.  $\chi$ undergoes rapid oscillations after inflation, but quickly settles down to 0. The time coordinate is the number of e-folds, and inflation ends at $N\approx 50$ for this choice of parameters.}
\label{fig:2}
\end{figure}

\begin{figure}[h!]
\centering
\includegraphics[width=0.7\textwidth]{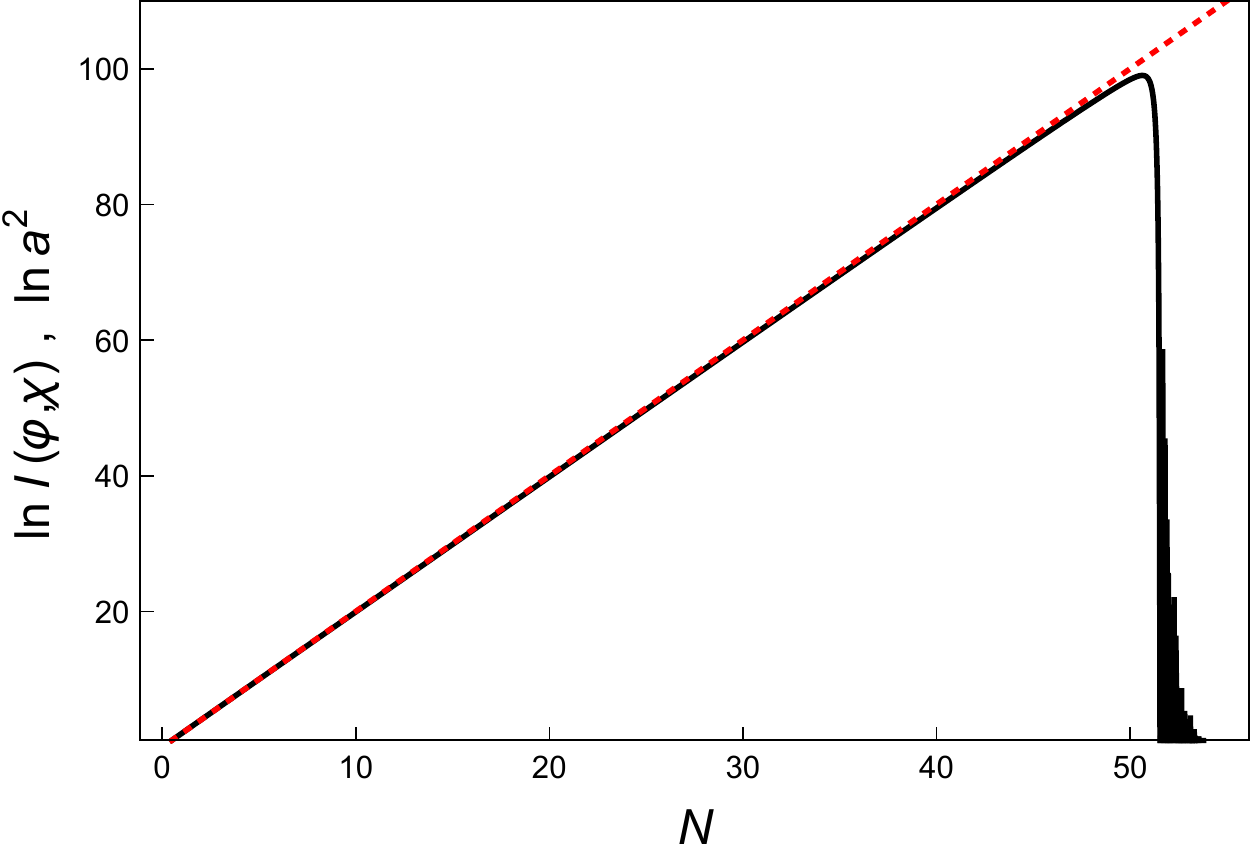}
\caption{
Evolution of $\bar I(\varphi,\chi)$ (black), compared to $a^2(t)$ (red, dotted). The black and red lines overlap for early times, indicating the behaviour $\bar I \propto a^2$ is well realized during inflation. The time coordinate is as in Figure \ref{fig:2}. }
\label{fig:fit}
\end{figure}

We can numerically solve for the evolution of the coupled $\{\varphi,\chi\}$ system during and after inflation, as well as for the evolution of $I(\varphi,\chi)$. This is shown in Figures \ref{fig:2} and \ref{fig:fit}, where the dynamics discussed thus far are confirmed. The parameters used are:
\bea
\label{params}
&&\lambda = 10^{-14} \;\;, \;\; g = 3 \times10^{-5} \;\;, \;\;m_{\chi} = 3 \times 10^{-6} M_{\rm Pl} \nonumber  \\
&& \chi_0 = 3 \times 10^{-2} \, M_{\rm Pl} \;\; , \;\; \varphi_0 = 20 \, M_{\rm Pl} \;\; , \;\; M = \frac{4 \chi_0 M_{\rm Pl}^2}{\varphi_0 ^2} = 3 \times 10^{-4} \, M_{\rm Pl},
\eea
where $\varphi_0$ is the initial value of $\varphi$. We also demonstrate that  $I(\varphi,\chi)$ scales as $a^2(t)$ in the slow-roll regime, before quickly jumping to unity. This is seen in Figure \ref{fig:fit}.
The resulting magnetic field power spectrum with the current set of parameters can be computed from \eqref{Bspecnow-Iinc}, which reads
\begin{equation}
\frac{d \big\langle \vec{B}^{\,2} \big\rangle}{d \ln k} (t_{\rm now})  \simeq \left( 10^{-11} \, \frac{{\rm e}^{-2 N_*}}{\bar{I}_*} \, {\rm G}  \right)^2 \simeq \left( 10^{-55} \, {\rm G} \right)^2 \; ,
\label{Bnow_concrete}
\end{equation}   
where the inflationary energy scale is $H_I \simeq \sqrt{\lambda} \, \varphi^2 / (\sqrt{3} \, M_{\rm Pl}) \simeq 10^{13} \, {\rm GeV}$, and ${\rm e}^{-2 N_*}/\bar{I}_* \simeq {\rm e}^{- \chi_0 / M} \simeq 10^{-44}$.
Thus, as in the general argument in Sec.~\ref{subsec:problem}, there is a negligible production of magnetic fields.

A subtlety of this mechanism is the possibility of the presence of a small number of e-folds of $\chi$-domination, which would slightly modify the predictions of $n_s$ and $r$ at fixed $N$. The number of e-folds of $\chi$-domination can be made small by tuning $\chi_0$ and $M$ to be small, as we have done.  We will not study this issue further,
since it would not change the conclusion of this paper.

A further subtlety of this mechanism is the effect of \emph{preheating}, which occurs after inflation, in the phase when the inflaton is oscillating. It is well known that for special values of $g$ and $\lambda$ there can be significant production of the perturbation $\delta \chi$ via a parametric resonance instability. It is therefore reasonable to ask if preheating can compensate for the suppression of the magnetic fields due to the jump in $I$. 
We investigate this possibility in Appendix \ref{app:preheating}, and demonstrate that such a compensation is not possible. Provided that the hierarchy $\vert \vec B \vert \gg \vert \vec E \vert$ is satisfied at the end of inflation, then preheating will not violate this and $\vert \vec B \vert \gg \vert \vec E \vert$ continues to be true during and after preheating. This results in $\vert \vec{B} \vert \propto \bar{I} / a^2$ at all times.

\section{Conclusion}
\label{sec:conclustion}

In this work we have studied primordial magnetogenesis via the coupling through the electromagnetic kinetic term,
\be
{\cal L}_{\rm EM} = - \frac{I^2}{4} \, F_{\mu \nu} F^{\mu \nu} ,
\ee
where $I^2$ is dynamical in the early universe and settles down to unity before, at latest, the time of BBN.
Since the pioneering work by Ratra \cite{Ratra:1991bn}, couplings of this type have been widely studied in the context of magnetogenesis in the very early universe, aiming for the explanation of the existence of magnetic fields in the extragalactic regions, suggested by blazar observations \cite{Neronov:1900zz, Tavecchio:2010mk, Dolag:2010ni, Essey:2010nd, Taylor:2011bn, Takahashi:2013lba, Finke:2013tyq, Chen:2014rsa}.
While the behavior of $I$ is {\it a priori} arbitrary, except that $I \to 1$ toward the present to recover the known electromagnetism, the case where $I$ {\it only} increases throughout its dynamical period inevitably falls into a strong coupling regime, as the effective coupling $e/I$ to charged particles becomes large for some time, and one loses perturbative control \cite{Demozzi:2009fu}.
For this reason, the case with a decreasing function of $I$ has been considered more extensively.
However, it is known that this leads to the overproduction of electric fields and that the constraints on them from backreactions and CMB observations severely limits the viability of this scenario.
The only known possible ways out of this issue are to fine-tune the inflationary scale to be very low, close to the BBN energy scale \cite{Ferreira:2013sqa, Ferreira:2014hma}, or to have a prolonged period of $I$ being dynamical even after inflation \cite{Kobayashi:2014sga, Fujita:2016qab}.

Given this rather tight situation in the existing studies of this class of models, we have presented a novel hybrid scenario, wherein $I$ increases during inflation, though keeping $I >1$, and rapidly decreases to settle at $I = 1$ afterwards. This is accomplished by promoting $I$ to a function of both the inflaton $\varphi$ and another scalar field $\chi$, so that the motion of $\varphi$ drives the increase of $I$ during inflation and that of $\chi$ is responsible for its post-inflationary dynamics.
In particular, the VEV of $\chi$ can be stabilized by its interaction to $\varphi$ during inflation, and once $\varphi$ starts oscillating and decays away, it is destabilized and rolls down to its true potential minimum, inducing the rapid change of $I$ after inflation.
Since in this case electric fields are always negligible compared to the magnetic, $\vert \vec{E} \vert \ll \vert \vec{B}\vert$, this mechanism appears to easily avoid both the backreaction and strong coupling problems.
The price to pay is, however, an unavoidable suppression in the amplitude of the scale-invariant magnetic fields, as we have demonstrated in Sec.~\ref{subsec:problem},
essentially ruling out this mechanism as a successful model of large-scale magnetogenesis.

This conclusion is quite general and solid. It is straightforward to reach this conclusion from the (modified) Maxwell equations \eqref{Maxwell}. Whenever electric fields are smaller in amplitude, by $\vert \vec{E} \vert \ll \left( aH/k \right) \vert \vec{B}\vert$ ($\vert \vec{E} \vert \ll \vert \vec{B} \vert$ suffices for superhorizon modes), the scale-invariant part of physical magnetic fields behave as $\vert \vec{B} \vert \propto I / a^2$ throughout its evolution history.
Since $I \gg 1$ at the end of inflation is necessary to avoid the strong coupling issue, {\it any} implementation of mechanisms to achieve $I \to 1$ at a later stage inevitably leads to a large suppression, in addition to the standard dilution $\propto 1/a^2$ due to the expansion.
In the case of $I \propto a^2$ during inflation, which produces scale-invariant magnetic spectra, this suppression completely cancels the gain from the inflationary production of magnetic fields.
Let us emphasize that this is true whenever $\vert \vec{E} \vert \ll \vert \vec{B}\vert$ for large-scale modes, and if this hierarchy is reversed, then the backreaction and CMB constraints on $\vec{E}$ known in the literature again take over.

This analysis, though not ultimately successful for explaining blazar observations, opens up several new directions for research. Here we have considered a two-field model of inflation, which however follows a single-field trajectory during inflation. One could instead consider the effect of more complicated field space trajectories, and the incorporation of this into other scenarios for the evolution of $I$. For example, one could use multifield dynamics to realize the decreasing-$I$ scenario of \cite{Fujita:2016qab}, which specified $I$ as a function of time and not fields. 

An orthogonal direction of research is to identify other observables where the destabilization of moduli can become important. In our analysis, the function $I$ was exponentially sensitive to the value of $\chi$. It would be interesting to study other scenarios where such a sensitivity exists or can be useful. Finally, it will be interesting to develop string theoretic scenarios where the moduli destabilization utilized here can be realized.

\acknowledgments

We would like to thank Shinji Mukohyama, Marco Peloso, Evangelos Sfakianakis, and Jun'ichi Yokoyama for useful discussions.
HBM was supported in part by an Iranian MSRT fellowship. EM is supported by the National Science and Engineering Research Council of Canada via a Post-Graduate Scholar Doctoral fellowship. 
RN is supported by the Natural Sciences and Engineering Research Council of Canada and by the Lorne Trottier Chair in Astrophysics and Cosmology at McGill. The research of RB is supported in part by
NSERC and the Canada Research Chair program.

\appendix

\section{Magnetic Fields during Preheating}
\label{app:preheating}

We have demonstrated in Sec.~\ref{subsec:jumpmodel} a mechanism to shift the kinetic function $I$ from a large value to unity after inflation. This transition is invoked by a non-inflaton scalar field $\chi$ that is coupled to the inflaton $\varphi$ through ${\cal L}_{\rm int} [\varphi, \chi] = - g^2 \varphi^2 ( \chi - \chi_0 )^2 / 2$ in \eqref{interaction}. Sec.~\eqref{subsec:jumpmodel} only considers a smooth transition due to the homogeneous motion of $\chi$. However, if some modes of the inhomogeneous perturbation of $\chi$ are in the resonance bands of the oscillation of the inflaton, then these modes undergo a preheating stage after inflation \cite{Traschen:1990sw, Dolgov:1989us, Kofman:1994rk, Shtanov:1994ce, Kofman:1997yn, Greene:1997fu} (for recent reviews on preheating, see \cite{Allahverdi:2010xz, Amin:2014eta}) and may potentially alter the evolution of the electromagnetic fields through the kinetic coupling $I^2(\varphi, \chi) F^2$.
In this appendix, we rule out this possibility and show that, even if preheating takes place, it does not help amplify the produced magnetic field to an observable level.
In short, preheating does not evade the no-go result for scale-invariant magnetic spectra in \eqref{subsec:problem}, which is generally valid as long as $\vert \vec{B} \vert \gg \vert \vec{E} \vert$. Below, we summarize that this is indeed the case.

For a concrete calculation, we consider the system described by the action \eqref{action} and take the inflaton potential to be $V(\varphi) = \lambda \varphi^4 /4$, as in Sec.~\ref{subsec:jumpmodel}, and as studied in \cite{Moghaddam:2014ksa}. In this case, the background space averaged over inflaton oscillations behaves as a radiation-dominated (RD) one during preheating, and the background inflaton field $\phi$ oscillates with an amplitude $\phi_p / a$ and a frequency $\sqrt{\lambda} \phi_p$, namely $\phi \simeq (\phi_p / a) \sin ( \sqrt{\lambda} \phi_p \tau + \theta_0 )$, where $\tau$ is the conformal time and $\theta_0$ is a phase. 
The equation of motion for the the mode function of $\chi$ perturbation $\delta\chi \equiv \chi - \langle \chi \rangle \cong \chi - \chi_0$ reads, in Fourier space,
\begin{equation}
\left( \partial_\tau^2 + k^2 + a^2 g^2 \phi^2 + a^2 U_{\chi\chi} - \frac{\partial_\tau^2 a}{a} \right) \left( a \delta\chi_k \right) = 0 \; ,
\label{EOM-chi}
\end{equation}
where $\phi(t) \equiv \langle \varphi \rangle$ is the homogeneous inflaton mode, and the subscript $\chi$ denotes the derivative with respect to it. During preheating, we have $\partial_\tau^2 a / a \simeq 0$ for a radiation-dominated background. An efficient preheating requires $U_{\chi\chi} \ll g^2 \phi^2$, which we thus impose. Using the solution for $\phi$, we have an effective equation of motion
\begin{equation}
\left[ \partial_\tau^2 + k^2 + g^2 \phi_p^2 \, \sin^2 \left( \sqrt{\lambda} \phi_p \tau + \theta_0 \right) \right] \left( a \delta\chi_k \right) \simeq 0 \; .
\label{EOM-chiRD}
\end{equation}
It is clear from \eqref{EOM-chiRD} that the quantity $a \delta\chi_k$ is insensitive to the background expansion. The equation is a Mathieu equation and there are instabilities for certain bands of values of $k$ \cite{mclachlan}.
The resulting amplification depends on the widths of these bands, and we are particularly interested in the values of $g^2 / \lambda$ that enhance large-scale modes $k \simeq 0$. Realizing this, we take $g^2/\lambda = 2$ as a fiducial value.
Solutions to \eqref{EOM-chiRD} are characterized by an exponential growth and are approximately given by
\be
\delta \chi_k \simeq {\cal A} \, \frac{H_{I} }{k^{3/2}} \, {\rm e}^{\mu_k x} \, \frac{\sin x}{x}
\label{dchi-preh}
\ee
where $H_I$ is the Hubble parameter during inflation, $\mu_k$ is the Floquet exponent, and $x\equiv \sqrt{\lambda} \phi_p \tau$ is a dimensionless time variable. 
The value of ${\cal A}$ is related to the amplitude of $\delta\chi_k$ at the end of inflation $x = 0$. If $\chi$ is light throughout inflation, then ${\cal A} = 1/\sqrt{2}$. On the other hand, as in the case we are concerned here, if $V(\varphi) \propto \varphi^4$ is valid near the end of inflation and $g^2 / \lambda =2$, then the value of $\phi$ changes smoothly, and the modes outside the Hubble radius become massive for the last few e-folds of inflation, giving ${\cal A} \sim 0.07$. Note that in this case \eqref{dchi-preh} is valid for the modes already outside the Hubble radius by the time $\chi$ becomes massive.
For the case $g^2/\lambda = 2$, the growth exponent is a constant on large scales $\mu_k \approx \mu = 0.25$. For this value of $\mu$, while the solution \eqref{dchi-preh} is dominated by the first resonance band, it allows the amplification for $k \simeq 0$ modes.

\subsection{Gauge Field Perturbations during Preheating}
\label{subapp:gauge}

Once preheating amplifies the $\chi$ perturbations (around $\chi = \chi_0$), the produced quanta of $\delta\chi$ can in turn enhance the gauge field. At leading order in perturbation theory, the equation of motion of the gauge field is given by \eqref{EOM-V}, now with $\bar I$ including the contribution from $\delta\chi$, namely,
\begin{equation}
\bar I \equiv \sqrt{\langle I^2 \rangle} = 
\exp \left( - \frac{\phi^2}{4 M_{\rm Pl}^2} + \frac{\chi_0}{M} \right)
\exp \left(\frac{\langle \delta\chi^2 \rangle}{M^2} \right) \; ,
\label{barI-app}
\end{equation}
where we  have used $\langle {\rm e}^X \rangle \equiv \sum_{n=0}^\infty \langle X^n \rangle / n! = \exp ( \langle X^2 \rangle /2 )$ for a Gaussian field $X$.

The inflationary solution for the gauge potential is given in \eqref{Vlam-sol}, whereas the equivalent, corresponding mode functions of the comoving magnetic and electric fields are in \eqref{BcEc-infsol}. For the evolution after inflation, it is more intuitive and instructive to look at the Maxwell equations \eqref{Maxwell2}, instead of the equation of motion \eqref{EOM-V}. Let us emphasize that solving the former is completely equivalent to solving the latter, given a set of initial conditions. Eq.~\eqref{Maxwell2} immediately gives
\begin{eqnarray}
B_P^c(\tau, k) & = & \frac{\bar I (\tau)}{\bar I (\tau_{\rm end})} \, B_P^c(\tau_{\rm end}, k) 
- P k \bar I (\tau) \int_{\tau_{\rm end}}^{\tau} d\tau' \, \frac{E_P^c(\tau', k)}{\bar I (\tau')} \; , 
\label{solBc-general}\\
E_P^c(\tau, k) & = & \frac{\bar I (\tau_{\rm end})}{\bar I (\tau)} \, E_P^c(\tau_{\rm end}, k) 
+ \frac{P k}{\bar I (\tau)} \int_{\tau_{\rm end}}^{\tau} d\tau' \, \bar I (\tau') B_P^c(\tau', k) \; ,
\label{solEc-general}
\end{eqnarray}
where the comoving fields $\vec B^c$ and $\vec E^c$ are defined in \eqref{BcEc}, and subscripts $P=\pm$ and ``end'' denote the photon circular polarization and the end of inflation, respectively. In the case of $\bar I \propto a^2$ during inflation, we have, from \eqref{BcEc-infsol}, $B_P^c(\tau_{\rm end}, k) \simeq 3 P a_{\rm end}^2 H_I^2 / (\sqrt{2} k^{3/2})$ and $E_P^c(\tau_{\rm end}, k) \simeq - a_{\rm end} H_I / \sqrt{2k}$.
In \eqref{solBc-general}, we see $\bar I (\tau_{\rm end} ) < \bar I (\tau' > \tau_{\rm end})$ during the preheating period, $E_P^c (\tau_{\rm end}, k) \ll B_P^c (\tau_{\rm end}, k)$ on superhorizon scales at the end of inflation, and $k/(a_{\rm end} H_I) \ll 1$ for large-scale modes, and therefore the second term in \eqref{solBc-general} is negligible compared to the first one. 
Then \eqref{solBc-general} and \eqref{solEc-general} become
\begin{eqnarray}
B_P^c(\tau, k) & \simeq & \frac{\bar I (\tau)}{\bar I (\tau_{\rm end})} \, B_P^c(\tau_{\rm end}, k) \; , 
\label{solBc-approx}\\
E_P^c(\tau, k) & \simeq & \frac{\bar I (\tau_{\rm end})}{\bar I (\tau)} \, E_P^c(\tau_{\rm end}, k) 
+ P \, \frac{k}{\sqrt{\lambda} \phi_p}  \,
\frac{\int_{x_{\rm end}}^{x} dx' \, \bar I^2 (x')}{\bar I^2 (\tau)}
\, \frac{\bar I (\tau)}{\bar I (\tau_{\rm end})} \, B_P^c(\tau_{\rm end}, k) \; ,
\label{solEc-approx}
\end{eqnarray}
where $x' \equiv \sqrt{\lambda} \phi_p \tau'$. Using \eqref{barI-app} with \eqref{dchi-preh}, one can explicitly verify that $\int_{x_{\rm end}}^{x} dx' \, \bar I^2 (x') / \bar I^2 (\tau) \lesssim {\cal O}(10)$, and therefore \eqref{solBc-approx} and \eqref{solEc-approx} tell us that at most $E_P^c / B_P^c \lesssim k / (a_{\rm end} H_I) \ll 1$ for the scales of our concern. This also justifies the aforementioned assumption to neglect the second term in \eqref{solBc-general}.

The important point is in \eqref{solBc-approx}, which tells that
\begin{equation}
B_P^c \propto \bar I
\end{equation}
holds despite the presence of the preheating period. Therefore, the calculations in Sec.~\ref{subsec:problem} still hold throughout this stage and, hence, throughout the evolution of the magnetic field, leaving unobservable strength \eqref{PB-now}.
This completes the no-go result in the scenario we consider.

\subsection{Backreaction and End of Preheating}

The discussion in App.~\ref{subapp:gauge} relies on the implicit assumption that the calculations can be done perturbatively. One might worry that it would be invalid if preheating lasted too long. In this sub-Appendix, we estimate the ending time of reheating and verify the validity of our result in App.~\ref{subapp:gauge}.

Preheating will continue until the background dynamics or oscillations of $\phi$ are disrupted by the growth of perturbations, or else when the growth of perturbations backreacts on the perturbation equations of motion and shuts off the resonance. The absence of backreaction in the Friedman equation requires,
\be
\label{cond1}
\rho_{\rm EM} \,, \, \rho_{\delta\chi} \,,\, \frac{g^2}{2} \phi^2 \langle \delta\chi^2 \rangle \ll 3 M_{p} ^2 H^2 \; ,
\ee
while the background $\phi$ equation of motion is unchanged if,
\be
\label{cond2}
g^2 \vert \phi \vert \langle \delta \chi^2 \rangle \;,\;  
\frac{1}{2} \big\vert \bar I \cdot \bar I_{,\varphi}  \langle F^2 \rangle \big\vert \; \ll  V_{, \varphi} \; ,
\ee
and negligible backreaction on the $\delta\chi$ equation of motion requires,
\be
\frac{1}{2} \big\vert \bar I \cdot \bar I_{,\chi}  F^2 \big\vert \; \ll 
g^2 \phi^2 \sqrt{\langle \delta\chi^2 \rangle} \; .
\label{cond3}
\ee
These are the conditions that involve the preheating of $\delta\chi$, and it would suffice in the case of massless $\chi$.
However, in order to manifest $\bar I \gg 1$ to $\bar I \to 1$ by the transition from the false minimum at $\chi = \chi_0$ to the true one at $\chi \to 0$, the value of $U_{\chi\chi}$ must depend on the value of $\chi$.
Thus, once the produced variance $\langle \delta\chi^2 \rangle$ makes $\chi$ reach a certain value $\chi_c$ in some Hubble patch such that $U_{\chi\chi}(\chi_c) > g^2 \phi^2$, then the production terminates in that patch. The time duration of efficient production is for
\begin{equation}
\sqrt{\langle \delta\chi^2 \rangle} \le \vert \chi_0 - \chi_c \vert \; ,
\label{cond4}
\end{equation}
and it terminates when the equality is met.

Each of these conditions can be used to estimate the value of $\langle \delta\chi^2 \rangle$ at which preheating ends, and then from \eqref{solBc-approx} the growth of $\vec{B}$ can be computed. 
One can show that the conditions \eqref{cond1} and \eqref{cond3} are not as competitive, and on the other hand, using the result in \eqref{solBc-approx} with $\bar I$ in \eqref{barI-app}, \eqref{cond2} and \eqref{cond4} are translated to the upper bound on $\langle \delta\chi^2 \rangle$
\begin{equation}
\frac{\langle \delta\chi^2 \rangle}{M^2} \ll \min\left[
10^6 \left( \frac{10^{-4} M_{\rm Pl}}{M} \right)^2 \, , \;
17 - \ln \frac{a}{a_{\rm end}} - \frac{1}{2} \ln \frac{N_{\rm prod}}{100} \, , \;
\left( \frac{\chi_0 - \chi_c}{M} \right)^2
\right] \; ,
\label{cond-on-dchi}
\end{equation}
respectively, assuming $\phi_{\rm end} \sim 10^{-1} M_{\rm Pl}$ and $H_I \sim 10^{-6} M_{\rm Pl}$.
In the case of $(\chi_0 - \chi_c) / M \lesssim {\cal O}(1)$, the last condition in \eqref{cond-on-dchi} is the most stringent one; otherwise, the middle condition dominates.
In either case, $\langle \delta\chi^2 \rangle / M^2 \ll {\cal O}(10)$ and so $\sqrt{\langle \delta\chi^2 \rangle} < \chi_0$ is respected. In this regime, we expect the analysis in App.~\ref{subapp:gauge} to be a valid approximation.


\end{document}